\documentclass[12pt]{article} 

\newcommand{\be}{\begin{equation}}
\newcommand{\ee}{\end{equation}}
\newcommand{\ea}{\end{eqnarray}}
\newcommand{\ba}{\begin{eqnarray}}

\def\[{\left\lbrack}
\def\]{\right\rbrack}

\def\({\left(}
\def\){\right)}

\usepackage{graphicx}

\begin{document}

\title{DeBroglie-Bohm interpretation of a Ho\v{r}ava-Lifshitz quantum cosmology model}

\author{G. Oliveira-Neto and L. G. Martins\\
Departamento de F\'{\i}sica, \\
Instituto de Ci\^{e}ncias Exatas, \\ 
Universidade Federal de Juiz de Fora,\\
CEP 36036-330 - Juiz de Fora, MG, Brazil.\\
gilneto@fisica.ufjf.br, laysamartinsymail@yahoo.com.br\\
\phantom{1}\\
G. A. Monerat\\
Departamento de Modelagem Computacional,\\
Instituto Polit\'{e}cnico do Rio de Janeiro,\\
Universidade do Estado do Rio de Janeiro,\\
Rua Bonfim, 25 - Vila Am\'{e}lia - Cep 28.625-570,\\
Nova Friburgo, RJ, Brazil.\\
monerat@uerj.br\\
\phantom{1}\\
E. V. Corr\^{e}a Silva\\
Departamento de Matem\'{a}tica, F\'{\i}sica e Computa\c{c}\~{a}o, \\
Faculdade de Tecnologia, \\
Universidade do Estado do Rio de Janeiro,\\
Rodovia Presidente Dutra, Km 298, P\'{o}lo
Industrial,\\
CEP 27537-000, Resende-RJ, Brazil.\\
evasquez@uerj.br\\
}





\maketitle

\begin{abstract}
In the present letter, we consider the {\it DeBroglie-Bohm} interpretation 
of a Ho\v{r}ava-Lifshitz quantum cosmology model in the presence of a radiation 
perfect fluid. We compute the Bohm's trajectories for the scale factor and
show that it never goes to zero. That result gives a strong indication that 
this model is free from singularities, at the quantum level. We also 
compute the quantum potential. That quantity helps understanding why the scale 
factor never vanishes.
\end{abstract}




Any quantized theory of the gravitational interaction when applied to cosmology will face an
important problem related to its interpretation. The well established {\it Copenhagen} interpretation of 
quantum mechanics cannot be applied to such theories because it is not possible to apply a statistical interpretation
to a system composed of the entire Universe. One cannot repeat the experiments for that system.
A very interesting interpretation of quantum mechanics, which in many aspects leads to the same
results as the {\it Copenhagen} interpretation and can be applied to a system composed of the
entire Universe is the {\it DeBroglie-Bohm} interpretation of quantum mechanics \cite{bohm}, 
\cite{holland}. That interpretation has been applied to several models of quantum cosmology
with great success 
\cite{acacio1}, \cite{nelson}, \cite{barbosa}, \cite{gil}, 
\cite{gil2}. In most of those models, the authors compute the scale factor trajectory
and shows that this quantity never vanishes. That result gives a strong indication that those models are
free from singularities, at the quantum level. Another important quantity introduced by the
{\it DeBroglie-Bohm} interpretation is the quantum potential ($Q$) \cite{bohm}, \cite{holland}. 
For those quantum cosmology models, the determination of $Q$ helps understanding why the scale factor never vanishes.


Some years ago a very interesting geometrical theory of gravity was introduced \cite{horava}. That theory, now called 
Ho\v{r}ava-Lifstz (HL), has an explicit asymmetry between space and time, which manifests through an anisotropic scaling 
between space and time. That property means that the Lorentz symmetry is broken, at least at high energies, where that 
asymmetry between space and time takes place. At low energies the HL theory tends to General Relativity (GR), recovering
the Lorentz symmetry. 
Due to the asymmetry between space and time present in his geometrical theory of gravity, Ho\v{r}ava decided to formulate it
using the Arnowitt-Deser-Misner (ADM) formalism, which splits the four dimensional spacetime in a time line plus three dimensional 
space \cite{misner}. In the ADM formalism the four dimensional metric $g_{\mu \nu}$ ($\mu, \nu = 1,2,3,4$) is decomposed in terms 
of the three dimensional metric $h_{i j}$ ($i, j = 1,2,3$), of spatial sections, the shift vector $N_i$ and the lapse function $N$. 
In general all those quantities depend on space and time. In his original work, Ho\v{r}ava considered the simplified assumption that 
$N$ should depend only on time \cite{horava}. This assumption became known as the {\it projectable condition}. 
The gravitational action of the HL theory was proposed such that the kinetic component was constructed separately from the potential one.
The kinetic component was motivated by the one coming from general relativity, written in terms of the extrinsic curvature tensor. 
The potential component must
depend only on the spatial metric and its spatial derivatives. As a geometrical theory of gravity the potential component of the HL theory
should be composed of scalar contractions of the Riemann tensor and its spatial derivatives. In his original paper \cite{horava}, Ho\v{r}ava
considered a simplification in order to reduce the number of possible terms contributing to the potential component of his theory. It is called
the {\it detailed balance condition}. 
The HL theory have 
been applied to cosmology and produced very interesting models 
\cite{bertolami}, \cite{paul}, \cite{kord}, \cite{pedram}.

In the present letter, we consider the {\it DeBroglie-Bohm} interpretation of a Ho\v{r}ava-Lifshitz quantum cosmology model in the presence of a radiation 
perfect fluid. The model have already been studied in Ref. \cite{kord}. There, the authors solved the corresponding Wheeler-DeWitt equation, found
the wavefunction and computed an approximate value of the scale factor expected value. Here, we compute the exact scale factor expected value and the Bohm's 
trajectories for the scale factor of that model and show that they never go to zero. That result gives a strong indication that the model is free from singularities, at the 
quantum level. We also compare that trajectory with the scale factor expected value. Finally, we compute the quantum potential
for that model. That quantity helps understanding why the scale factor never vanishes.


Following the authors of Ref. \cite{kord}, we study, here, Friedmann-Robertson-Walker quantum cosmology models in the framework of a projectable HL gravity 
without detailed balance condition. The matter content of the models is a perfect fluid with equation of state: $p = \omega \rho$, where $p$ is the fluid
pressure and $\rho$ its energy density. The constant curvature of the spatial sections may be positive ($k=1$), negative ($k=-1$) or zero ($k=0$). The model
may be written in its Hamiltonian form with the aid of the ADM formalism \cite{misner} and the Schutz's variational formalism \cite{schutz}, \cite{schutz1}.
The authors of Ref. \cite{kord} did that and found the following total hamiltonian,
\be
\label{2}
H=N\left(-\frac{P_{a}^{2}}{4a}-g_{c}ka+g_{\Lambda}a^{3}+\frac{g_{r}k^{2}}{a}+\frac{g_{s}k}{a^{3}}+\frac{P_{T}}{a^{3\omega}}\right).
\ee
The phase space of those models are described by the scale factor $a$, a variable associated to the fluid $T$ and their canonically conjugated momenta
$P_a$ and $P_T$, respectively. All of them are functions only of the time coordinate $t$. The coefficients $g_c$, $g_{\Lambda}$, $g_r$ and $g_s$, are 
coupling constants introduced by the Ho\v{r}ava-Lifshitz theory \cite{maeda}. The authors of Ref. \cite{kord}, restricted their attention to the gauge $N=a^{3\omega}$.

The quantum version of the theory is obtained by replacing the moments by their corresponding operator expressions and the application of the resulting total
hamiltonian operator to a wavefunction $\psi(a,T)$. The resulting equation is the Wheeler-DeWitt equation. For the present model it is given by \cite{kord},
\begin{eqnarray}
 \left[\frac{1}{a}\frac{\partial^{2}}{\partial a^{2}}-\frac{p}{a^{2}}\frac{\partial}{\partial a}-\frac{4i}{a^{3\omega}}\frac{\partial}{\partial T}+4\left(-g_{c}ka+g_{\Lambda}a^{3}+\frac{g_{r}k^{2}}{a}+\frac{g_{s}k}{a^{3}}\right)\right]\psi(a,T)=0, 
 \label{3}
\end{eqnarray}
where $p$ is a parameter that accounts for the ambiguity in the ordering of factors $a$ and $P_a$ in the first term of eq. (\ref{2}). The authors of Ref. \cite{kord},
restricted their attention to the case $p=1$. Now, we perform the following separation of variables in $\psi(a,T)$,
\begin{eqnarray}
\label{4}
 \psi(a,T)=e^{-iET}\psi(a),
\end{eqnarray}
Introducing Eq. (\ref{4}) in eq. (\ref{3}), we find the following equation for $\psi(a)$,
\begin{eqnarray}
\label{5}
  \left[a^{2}\frac{d^{2}}{da^{2}}-a\frac{d}{da}+4\left(-g_{c}ka^{4}+g_{\Lambda}a^{6}+g_{r}k^{2}a^{2}+g_{s}k+Ea^{3-3\omega}\right)\right]\psi(a) = 0.
\end{eqnarray}
For $k\neq 0$, the authors of Ref. \cite{kord}, found exact solutions to a simplified version of eq. (\ref{5}), obtained by setting $g_c = g_\Lambda = g_r =0$.
The motivation for doing that simplification is the fact that one may neglect the terms with those coefficients at the beginning of the Universe.
In particular, let us study the solution to the model with radiation ($\omega=1/3$) using the {\it DeBroglie-Bohm} interpretation.

\subparagraph{Solution for $k\neq0$ and $\omega=1/3$}

For the present case, the authors of Ref. \cite{kord} found the following solution to a simplified version of eq. (\ref{5}), obtained by setting $g_c = g_\Lambda = g_r =0$,
\begin{equation}
\label{18}
\psi_E(a)=a J_{\sqrt{1-4kg_s}}\left(2\sqrt{E}a\right),
\end{equation}
where $J$ is the Bessel function of the first kind. Now, introducing $\psi_E(a)$ in eq. (\ref{4}), 
it is possible to construct a wave-packet out of the resulting expression. 
In order to do that, one has to integrate, the resulting expression, over $E$, from 0 to $\infty$,
with an appropriate weight function. The authors of Ref. \cite{kord} did that and found the following wave-packet,
\begin{equation}
\label{19}
\psi(a,T) = \frac{a^{1+\sqrt{1-4kg_{s}}}}{(\gamma-iT)^{1+\sqrt{1-4kg_{s}}}}e^{-\frac{a^{2}}{\gamma-iT}},
\end{equation}
where $\gamma$ is a small positive constant. For simplicity let us choose $\gamma=1$. We want, now, to write $\psi(a,T)$ 
eq. (\ref{19}) in its polar form,
\begin{equation}
\label{7,5}
\psi(a,T)=R(a,T)e^{iS(a,T)}. 
\end{equation}
After some calculation, we obtain,
\begin{eqnarray}
\label{20}
\psi(a,T)&=&\frac{a^{1+\sqrt{1-4kg_{s}}}}{(1+T^{2})^{\frac{1}{2}+\frac{\sqrt{1-4kg_{s}}}{2}}}e^{-\frac{a^{2}}{(1+T^{2})}}\\\nonumber
&\times&e^{i\left[\left(1+\sqrt{1-4kg_{s}}\right)\tan^{-1}(T)-\frac{Ta^{2}}{(1+T^{2})}\right]}.
\end{eqnarray}
From eq. (\ref{20}), we may identify $R(a,T)$ and $S(a,T)$ as,
\begin{eqnarray}
\label{21}
 R(a,T) &=&\frac{a^{1+\sqrt{1-4kg_{s}}}}{(1+T^{2})^{\frac{1}{2}+\frac{\sqrt{1-4kg_{s}}}{2}}}e^{-\frac{a^{2}}{(1+T^{2})}} , \\
\label{21,5}
 S(a,T) &=& \left(1+\sqrt{1-4kg_{s}}\right)\tan^{-1}(T)-\frac{Ta^{2}}{(1+T^{2})}.
\end{eqnarray}

From $\psi(a,T)$ eq. (\ref{20}), it is possible to compute the scale factor expected value, using the expression,
\begin{equation}
\label{10}
\langle a\rangle (T) = \frac{\int_{0}^{\infty}\psi(a,T)a\psi^{*}(a,T)da}{\int_{0}^{\infty}\psi(a,T)\psi^{*}(a,T)da}.
\end{equation}
Therefore, introducing $\psi(a,T)$ eq. (\ref{20}) in eq. (\ref{10}), we obtain the following exact expression for the scale factor expected value,
\begin{equation}
\label{22}
\langle a\rangle (T) = \sqrt{1+T^{2}}\left(\frac{1-4kg_s+\sqrt{1-4kg_{s}}}{\sqrt{2}}\right)\frac{\Gamma\left(\sqrt{1-4kg_{s}}\right)}{\Gamma\left(\frac{3}{2}+\sqrt{1-4kg_{s}}\right)}.
\end{equation}
The authors of Ref. \cite{kord} found an approximate expression for the scale factor expected value.

Now, we want to describe that model in the {\it DeBroglie-Bohm} interpretation. In that interpretation, we may compute the scale factor trajectory
using the following equation \cite{holland},
\begin{equation}
\label{12}
P_{a}=\frac{\partial S(a,T)}{\partial a},
\end{equation}
where $S(a,T)$ is the phase, eq. (\ref{21,5}), of $\psi(a,T)$, eq. (\ref{20}). From Ref. \cite{kord}, we have that,
\be
\label{13}
P_{a}=-2\frac{a\dot{a}}{N}.
\ee
In the gauge $N=a^{3\omega}$, for the present case of radiation where $\omega=1/3$, and $P_a$ eq. (\ref{13}), we may write eq. (\ref{12}) as,
\be
\label{23}
 \dot{a} = -\frac{1}{2}\frac{\partial S(a,T)}{\partial a}.
\ee
Introducing the phase $S(a,T)$ eq. (\ref{21,5}) in eq. (\ref{23}) and performing the partial derivative and integration indicated, we obtain the following scale factor trajectory,
\be
\label{24}
a(T)=a_{0}\sqrt{\frac{1+T^{2}}{1+T_{0}^{2}}},
\ee
where $T_0$ is the initial value of $T$ and $a_0$ is the value of $a(T_0)$. Comparing the scale factor expected value 
eq. (\ref{22}) and its {\it DeBroglie-Bohm} trajectory eq. (\ref{24}), we notice that those expressions depend on $T$ in the
same way. If we fix $T_0=0$ and
\be
\label{25}
a_{0} = \left(\frac{1-4kg_s+\sqrt{1-4kg_{s}}}{\sqrt{2}}\right)\frac{\Gamma\left(\sqrt{1-4kg_{s}}\right)}{\Gamma\left(\frac{3}{2}+\sqrt{1-4kg_{s}}\right)},
\ee
both scale factor expected value eq. (\ref{22}) and its {\it DeBroglie-Bohm} trajectory eq. (\ref{24}) are the same.
That trajectory eq. (\ref{24}), represents an universe that starts, at $T=0$, from a minimum size $a_0$ eq. (\ref{25}) and expands to an infinity size, as $T$, when $T\to \infty$.
Therefore, it is free from singularities at the quantum level.
Observing the expressions of $\langle a\rangle$ eq. (\ref{22}) or $a(T)$ eq. (\ref{24}), which are the same for
$a_0$ given by eq. (\ref{25}), we notice that they depend on two parameters: $k$ and $g_s$. After a detailed numerical study on how $\langle a\rangle$ or $a(T)$ 
depend on those parameters, we reached the following conclusions. Since $k$ and $g_s$ always appear together, as $-4kg_s$, in the expressions of $\langle a\rangle$ or 
$a(T)$, we have three different cases: (i) $-4kg_s<0$, (ii) $-4kg_s=0$ and (iii) $-4kg_s>0$. It is important to notice that for case (i) $\langle a\rangle$ or $a(T)$ 
will be real-valued only if the condition $1-4kg_s>0$ is satisfied. That restriction, in the values of the parameter $g_s$, for case (i), shows the limitations of the 
present simplification used in order to obtain solution (\ref{18}). We hope, in a future work, solving the complete equation (\ref{5}). If one fixes the value of $g_s$ 
such that $-4kg_s$ have the same absolute value, for cases (i) and (iii), one obtains that $\langle a\rangle$ or $a(T)$ always have the greater $a_0$ and expand more 
rapidly for case (iii), next for case (ii) and finally for case (i). Those results are in agreement to the corresponding classical model \cite{kord}.
An example comparing the scale factor expected values or the {\it DeBroglie-Bohm} trajectories for those three cases is shown in Figure 1.

Using the quantum potential $Q(a,T)$ it is possible to understand why the scale factor trajectory eq. (\ref{24}) does not go to
zero when $T\to 0$. In order to obtain the quantum potential $Q(a,T)$, first we introduce the polar form of the wave function eq. (\ref{7,5}) in the Wheeler-DeWitt equation (\ref{3}),
for the radiation perfect fluid ($\omega=1/3$) and the factor ordering ambiguities parameter $p=1$. It will produce two equations: a real one and a purely imaginary one. The real equation is given by,
\begin{eqnarray}
\label{25,1}
 -4\left[\frac{\partial S(a,T)}{\partial T}\right] + \left(\frac{\partial S(a,T)}{\partial a}\right)^{2}+Q(a,T)
 + V(a,T) = 0, 
\end{eqnarray}
\noindent where the quantum potential $Q(a,T)$ is,
\begin{eqnarray}
\label{25,2}
  Q(a,T)=-\frac{1}{R(a,T)}\frac{\partial^{2} R(a,T)}{\partial a^{2}} + \frac{1}{a}\frac{1}{R(a,T)}\frac{\partial R(a,T)}{\partial a}.
\end{eqnarray}
The second part of the expression of $Q(a,T)$, comes from the choice of taking in account factor order ambiguities \cite{nelson1}, \cite{dongshan}. 
For the simplified version of the Wheeler-DeWitt equation considered, in order to obtain the solution eq. (\ref{18}), the classical potential $V(a,T)$ is,
\begin{eqnarray}
\label{25,3}
 V(a,T)=\frac{-4kg_{s}}{a^{2}}.
\end{eqnarray}
In order to compute $Q(a,T)$, we introduce $R(a,T)$ eq. (\ref{21}) in eq. (\ref{25,2}) and obtain the following expression,
\be
\label{26}
Q(a,T) = \frac{4kg_{s}}{a^2}+\frac{4(1+\sqrt{1-4kg_{s}})}{1+T^2}-\frac{4a^2}{(1+T^2)^2}.
\ee
Now, we have to evaluate the quantum potential eq. (\ref{26}) over the {\it DeBroglie-Bohm} trajectory eq. (\ref{24}) with $a_0$ given by eq. (\ref{25}) and setting $T_0=0$. We may write the
quantum potential as a function of $T$ ($Q(T)$),
\ba
\label{27}
Q(T) = \left[\frac{4kg_s}{a_0^2}+4(1+\sqrt{1-4kg_{s}})-4a_0^2\right] \frac{1}{1+T^2}.
\ea

After a detailed numerical study of $Q(T)$ eq. (\ref{27}), we find that for the three cases considered above: (i) $-4kg_s<0$, (ii) $-4kg_s=0$ and (iii) $-4kg_s>0$, $Q(T)$ is positive and finite at $T=0$,
then it decreases as $T$ increases and asymptotically it goes to zero when $T \to \infty$, for all possible values of $g_s$ (depending on each case). 
An example of the quantum potential, over the {\it DeBroglie-Bohm} trajectory Q(T), for the three situations described in the text is shown in Figure 2.
It is interesting to notice that, for case (i)
$-4kg_s<0$, observing the classical potential $V(a,T)$ eq. (\ref{25,3}), we notice that $V(a,T)$ gives rise to a potential well when $T\to 0$. The effective potential, which is given by the sum of 
$V(a,T)$ eq. (\ref{25,3}) with $Q(a,T)$ eq. (\ref{27}), is positive and finite at $T=0$, then it decreases as $T$ increases and asymptotically it goes to zero when $T \to \infty$, for all possible 
values of $g_s$. For case (ii) $-4kg_s=0$, the classical potential $V(a,T)$ eq. (\ref{25,3}) is zero. For that case, the model becomes similar to the corresponding one in quantum cosmology based on 
general relativity coupled to a radiation perfect fluid. The {\it DeBroglie-Bohm} interpretation of that model has already been treated in Ref. \cite{nivaldo} and their results are in agreement with 
ours. For case (iii) $-4kg_s>0$, observing the classical potential $V(a,T)$ eq. (\ref{25,3}), we notice that $V(a,T)$, along with $Q(a,T)$, also produces a potential barrier that prevents the value 
of the scale factor ever to go through zero, at $T=0$. Therefore, for the three cases there will be a potential barrier, at $T=0$, that prevents the value of the scale factor ever to go through zero.

\begin{figure}
\includegraphics[scale=0.4]{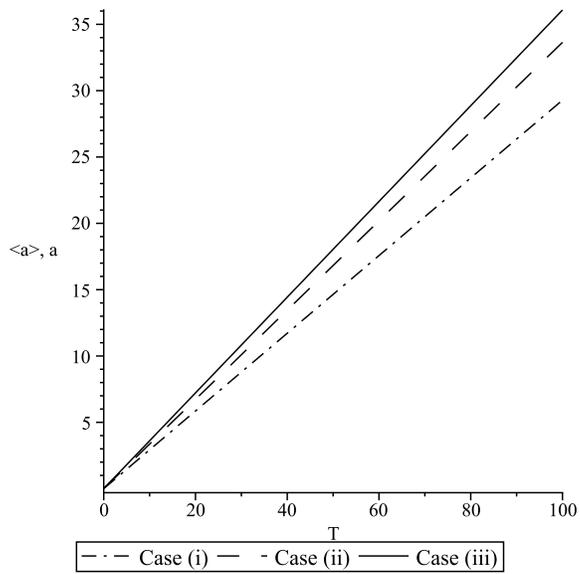}
\caption{Scale factor expected values or the {\it DeBroglie-Bohm} trajectories for the three cases described in the text with $T_{0}=0$. For (i) $k=1$, $g_s=1/5$; 
(ii) $k=0$, $g_s=1/5$; (iii) $k=-1$, $g_s=1/5$. 
}
\label{figure1}
\end{figure}

\begin{figure}
\includegraphics[scale=0.4]{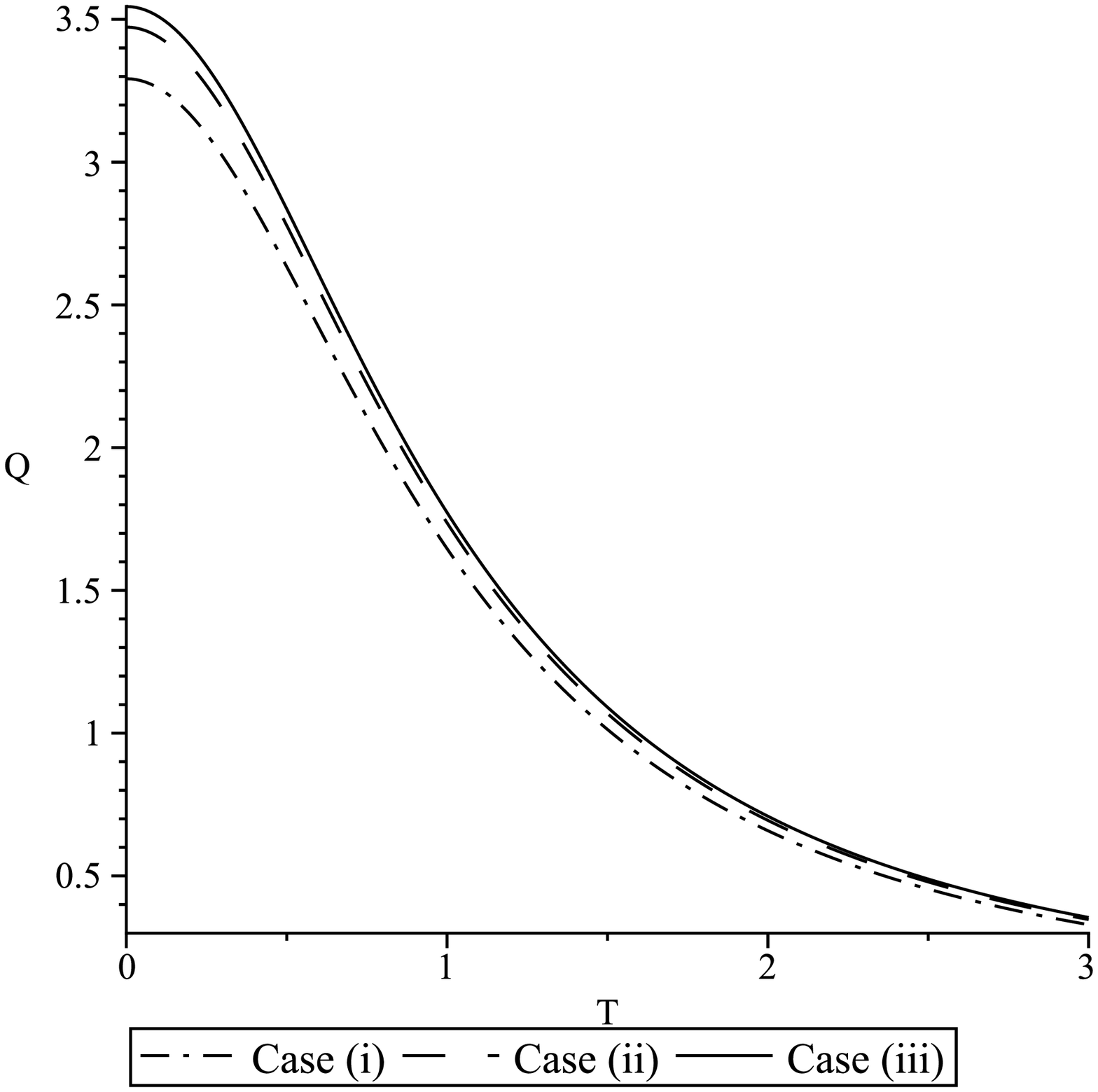}
\caption{Quantum potential over the {\it DeBroglie-Bohm} trajectory Q(T) for the three cases described in the text with $T_{0}=0$. For (i) $k=1$, $g_s=1/5$; 
(ii) $k=0$, $g_s=1/5$; (iii) $k=-1$, $g_s=1/5$. 
}
\label{figure2}
\end{figure}



{\bf Acknowledgements.} 
L. G. Martins thanks CAPES for her scholarship.
G. A. Monerat thanks UERJ for the Proci\^{e}ncia grant, via FAPERJ.

\end{document}